\documentclass[journal]{IEEEtran}

\ifCLASSINFOpdf
 
\else
 
\fi

\usepackage{enumerate}
\usepackage{algorithm}
\usepackage{algpseudocode}
\usepackage{graphicx} 
\graphicspath{ {images/} }
\usepackage[rightcaption]{sidecap}
\usepackage{wrapfig}
\usepackage{tabu}
\usepackage{amssymb}
\usepackage{hyperref}
\begin{document}

\title{ Partitioning Distribution Networks for Integrated Electrification Planning}

\author{Olamide Oladeji,~\IEEEmembership{Member,~IEEE,}
        Pedro Ciller Cutillas,
        Fernando de Cuadra,
        and~Ignacio Perez-Arriaga,~\IEEEmembership{Life~Fellow,~IEEE}}

\markboth{Journal Paper Pre-print August 2018}%
{Shell \MakeLowercase{\textit{et al.}}: Bare Demo of IEEEtran.cls for IEEE Journals}

\maketitle

\begin{abstract}
In many developing countries, access to electricity remains a significant challenge. Electrification planners in these countries often have to make important decisions on the mode of electrification and the planning of electrical networks for those without access, while under resource constraints. An integrated approach to electrification planning in which traditional grid electrification is complemented off-the-grid technologies such as off-grid microgrids and stand-alone systems can enable the economic provision of electricity access in these regions. This integrated planning approach can be facilitated by determining the least-cost mode of electrification - i.e by electric grid extension  or off-grid systems -  for non-electrified consumers in a region under analysis, while considering technical, economic and environmental constraints. Computational clustering methods the identification of consumer clusters (either as clusters of off-grid microgrids, stand-alone systems or grid-extension projects) can be undertaken using computational clustering methods. This paper presents a novel computational approach to achieve this purpose. This methodology involves exploiting the grid network that connects all consumers, by greedily partitioning the network to identify clusters of consumers to be electrified by grid-extension and off-grid microgrid systems. Using test cases and sensitivity analyses, we implement and benchmark this top-down approach with those obtained from a bottom-up clustering methodology used by the Reference Electrification Model, a model obtainable in literature. Results presented show that the alternative top-down methodology proposed can compare favorably, in terms of global electrification costs, with a bottom-up approach to rural electrification planning.
\end{abstract}


\IEEEpeerreviewmaketitle

\section{Introduction}

\IEEEPARstart{I}{n} many parts of the world, access to basic electricity services remains a significant challenge. In fact, in 2015, the International Energy Agency (IEA) estimated that about 1.1 billion lacked access to electricity. \cite{international_energy_agency_weo-2017_2017}. Electrification planners in countries where energy access issues persist can incorporate multiple technological alternatives such as grid-extension, off-grid microgrids and stand-alone home systems in order to undertake optimal electrification planning for those without access subject to regulatory, economic and political constraints. These planners have to make decisions on what mode of electrification, how to electrify or design any networks, for millions of people while aligning with national electricity policies and the universal service obligation to provide electricity to consumers on demand.

If the consumers in a region are mandated to be connected to the grid, then the planning involved is one of designing the substations and associated grid distribution network. This distribution planning can be undertaken via various planning and network design tools such as the Reference Network Model presented in \cite{domingo_reference_2011}. While having all new consumers fully connected to the grid may be a desirable outcome, it may not be the least-cost electrification strategy. For example, off-grid microgrids may be the least expensive mode of electrification in areas where there are dense clusters of consumers at significant distances from the existing grid. The natural geospatial spread of consumers into villages may be such that remote natural clusters or 'villages' exist for which grid extension may be prohibitively expensive. For any such designated off-grid locations, we can design microgrid generation systems and their associated distribution networks to provide electricity for connected consumers. A consumer may also exist as a stand-alone candidate, with its own dedicated power system. The design of such off-grid (microgrid) systems typically involves the determination of the optimal mix of energy sources and is a well-studied and researched problem in literature \cite{li_local_2016}\cite{homerBr_2016}. Similary, the design of the microgrid distribution network can be undertaken via the same well-studied approaches for grid extension. In all cases discussed, i.e. grid extension, off-grid microgrid and stand-alone systems, the costs of the electrification planning can be estimated as part of the system design process.
\\
It has been highlighted that cost is a significant factor serving as a barrier to achieving global energy as it informs rural electrification investment \cite{international_energy_agency_weo-2017_2017}. This has several implications, given rise to the following questions; how can we take advantage of the characteristics of a consumers in a given region, such as their geospatial location and expected demand, to find the least-cost electrification mode? Can integrated combinations of both the grid-extension and off-grid electrification planning lead to overall costs than when using solely one mode of electrification?  If so, how can we systematically determine these combinations and undertake integrated electrification planning?

To this end, many software tools and computer models have been proposed to aid these challenging integrated electrification planning efforts for energy access, allowing the consideration of the dual electrification strategies of extending the grid and providing off-grid systems.  Computational techniques can provide enabling data-driven platforms to analyze these planning efforts and to support data-driven decisions for large-scale electrification projects. For instance, the planner can utilize software tools to understand the most economical way to electrify a large amount of consumers in unserved or underserved regions, to simulate network growth and understand reinforcements required, to design optimal networks of both grid and off-grid systems, to understand the trade-offs between energy sources and undertake optimal generation system design among others. Recent advancements in computational processing capabilities and techniques mean that many of these decisions can now be made more data-driven and undertaken for large-scale regions, often involving millions of consumers and within relatively short time periods.

Methodology-wise, existing tools for integrated rural planning incorporate varying heuristics in determining which consumers are designated off-grid or grid-extension. Many of these approaches focus on existing natural clusters such as communities or villages and utilize heuristics involving thresholds such as the distance to the nearest grid line in order to designate off-grid and grid-extension consumer candidates. While the use of such heuristics allows for quick decision-making when undertaking large-scale integrated electrification planning, they ignore important factors that would otherwise lead to more robust planning.

We therefore, begin the next section by reviewing some of the widely-used tools and models for integrated rural electrification planning, focusing on the different approaches to electrification mode recommendation. After discussing these tools and highlighting some limitations in their approaches, we introduce the Reference Electrification Model (REM) and its rigorous bottom-up approach to the rural electrification planning problem. In section IV., we then discuss the details of an alternative top-down approach to REM's underlying bottom-up clustering and planning process, and compare the performance of these differing approaches on an integrated planning test case. The latter sections of this paper review the results of this comparative analyses and the paper concludes with a discussion on insights garnered.

\section{Support Systems for Electrification Planning}
 In general, rural electrification planning for energy access problem can be seen a large-scale optimization problem of assigning consumers to electrification modes while minimizing overall investment or system cost. Unlike traditional urban electrification planning in which electrification planning primarily involves just planning the grid distribution network, rural electrification planning, in particular for energy access, may involve a combination of extending the grid and designing off-grid microgrid systems as the least-cost solution and these have to be taken into account in the planning process. In this section, we identify a number of tools which support these rural electrification planning decisions and discuss their varying approaches to this problem. 
The Network Planner discussed in \cite{kemausuor_electrification_2014}  is one such tool for geospatial rural electrification planning which has been well-documented in literature. It determines if grid extension is favorable in comparison to microgrids using a modified version of Kruskal's shortest path algorithm, finding the shortest paths between potential grid-connected locations. It limitations include a low-level granularity in its final recommendation as demand and the final electrification mode recommendation is analyzed at a community cluster level rather than building level. It also does not properly capture other costs such as those due to network reinforcements in its underlying computation methodology. Finally, there is a lack of detailed network design in the final recommended output. Despite these, Network Planner has been applied to many real world electrification planning projects. For instance, \cite{kemausuor_electrification_2014} shows an application of the Network Planner to rural electrification planning in Ghana.

\cite{fronius_rural_2001} describes another GIS-based rural electrification planning tool called LAPER developed by the EDF. LAPER receives input data on network, geographical limits and an initial state design for a given community of villages before then proceeding to determine - via step-by-step series of replacement stages using alternative energy sources and configurations - how to connect as many villages as possible under the given geographical constraints. Its optimization algorithm seeks to optimize the global cost of electrification over the entire community of villages i.e. the sum of the investment and operation and maintenance costs. It incorporates multiple criteria such as political, financial and development objectives in its decision making. The final output of LAPER is a GIS representation of the electrification mode.  A full description of LAPER is provided in \cite{fronius_rural_2001}. 

IntiGIS, described in \cite{dominguez_gis_2009}, approaches GIS-based decision-support for electrification planning in a different way to the above. First, IntiGIS estimates the `Levelized Electric Cost' (LEC) of competing technologies (grid, PV, wind, etc.) for each community location provided before outputting the most competitive technology for that location based on LEC calculations. IntiGIS then provides this information on a GIS visualization output and does not provide any network designs \cite{amador_application_2005} \cite{amador_spatial_2006}.

GEOSIM is a commercial rural electrification decision support software developed by Innovation Energie Development \cite{innovation_energie_development_ied_geosim_2017}. Like Network Planner, GEOSIM also incorporates GIS into its approach to least-cost electrification mode recommendation. Rather than consumer-level designs and analyses, GEOSIM finds `development centers' within communities using a gravity probability model and determines the least-cost electrification mode to supply each of these centers. 

GEOSIM takes in a number of technical and economic inputs and also allows for users to receive outputs on details such as ``the percentage of people living in an electrified settlement". In addition, it also indicates the location of isolated settlements - settlements that are too far from electrified development centers and it estimates investment plans to provide power for basic social amenities (school, hospital) in such isolated settlements \cite{innovation_energie_development_ied_geosim_2017} \cite{rambaud-measson_introduction_2007}.
\\
In addition to \cite{innovation_energie_development_ied_geosim_2017}, most of the other tools available in literature focus on the applications of simple heuristics in determining electrification planning. For example, the authors of \cite{bertheau_electrification_2016}, utilized heuristics on the geospatial features such as distance and population density to recommend regions for off-grid and grid electrification. In \cite{mentis_benefits_2016}, another GIS-based rural electrification model is presented and applied to electrification planning in Ethiopia. The model takes in GIS inputs such as proximity to grid as well as resources data (solar, wind potential and mining reserves) and evaluates locational Levelized Cost of Electricity (LCOE) values in making visual recommendations of least-cost mode of electrification.  An extensive review of some of these tools can be found in \cite{cader_electrification_2016}. 

Overall, we can identify the following limitations in these tools:

\begin{itemize}
\item	The lack of analyses of electrification mode on a building-by-building or consumer-level basis.
\item The consumer groups or `clusters' for analysis for electrification are not computationally-determined least-cost clusters of consumers. The tools above largely examined consumer groups for electrification recommendation based on pre-defined natural clusters such as villages or ``development centers". The limitation of this is that there may be other possible groupings of consumers for service by microgrids or grid-extension which may have lower costs per off-grid generation or grid-extension project. 
\item	The lack of incorporation of technical and geospatial network designs alongside recommendation mode outputs. By not considering grid network design topologies, potential cost-savings opportunities associated with the grid are not captured.
\item The final output of these tools do not include the final electrification design plans, making it difficult to compare planning costs.
\end{itemize}
 
The Reference Electrification Model described in \cite{Amatya_WP_2018}  seeks to overcome the aforementioned limitations of existing planning tools for electrification planning for energy access by providing more granularity in its approach to electrification mode recommendation, such that the least-cost electrification strategy are determined at individual building level. In the next section, we discuss REM's underlying approach to integrated electrification planning and introduce the alternative network partitioning procedure which is the focus of this paper. 

\section{The Reference Electrification Model}  
REM uses a bottom-up clustering algorithm to identify either off-grid microgrid clusters or grid-extensions as the least-cost electrification option while incorporating both techno-economic and geospatial information.
 REM is also able to both provide final network designs for all recommended systems and better quantify network costs in its underlying recommendation and clustering processes. \cite{Amatya_WP_2018} provides an extensive description of the 'bottom-up', agglomerative clustering methodology, showing the performance of the clustering and resulting large-scale planning solutions. The approach is referred to as ``bottom-up" since it begins with the assumption that each consumer is in its own cluster before systematically merging consumers based on least cost electrification modes.  
 \\
 Under this bottom-up approach, a Delaunay Triangulation (DT) connecting every consumer of an analysis region is built. Afterwards, arcs of the Delaunay are sorted in increasing order of length and evaluated to determine if the two clusters located at its ends should be joined in one cluster. The assumption here is that initially all edges (connections) are not activated and so every customer node is in its own cluster of only that node. Agglomeration occurs when edges are activated - based on defined parameter comparisons - such that customers at both node ends of an edge are connected into same cluster.  
The algorithm loops several times until no new connection is activated and at the end of this process, the off-grid clusters would have been calculated. The idea of merging is such that the savings of being together compensate the extra connection costs. To obtain grid-extension clusters, the inactive arcs of the DT are reevaluated, now comparing the cost/savings balance of being connected to the grid together against being electrified separately (with at least one of them connected to the grid). Depending on the most inexpensive configuration for merging, the clusters are joined and the algorithm loops several times until no new connection is activated. At the end of this process, the on-grid clusters have been calculated. 
Utilizing these clusters, REM then undertakes the individual systems design, designing the network and generation systems for off-grid systems as well as the entire grid-extension network for on-grid systems. In addition to the final system designs, REM also estimates the overall annualized system costs. For integrated planning, this would correspond to the summation of the offgrid annuities and the grid-extension annuities. 
In particular, REM utilizes two types of costs in its optimization and final system cost estimation procedures; direct monetary costs and indirect societal costs. For direct monetary costs, REM considers the annuitized investment cost or CAPEX, annual operation and maintenance (O \& M) and management costs and the direct energy cost (\$ per kWh). The indirect cost is incorporated through the Cost of Non-Served Energy(CNSE), a cost that penalizes every unit of energy not supplied when demanded.  These costs are incorporated into the optimization, clustering and system design procedures in the model and also feed into the final estimated system cost for an electrification planning exercise.

Examples of applications of REM have been described in \cite{li_local_2016} \cite{ellman_reference_2015} \cite{cotterman_enhanced_2017} and \cite{drouin_geospatial_2018}. 
 In \cite{cotterman_enhanced_2017}, improvements on the underlying methods are described with additional applications of REM for regions in Rwanda and Uganda. \cite{cotterman_enhanced_2017} also addresses methods for the estimation of electrification status and for the quantification of upstream reinforcement. \cite{drouin_geospatial_2018} focuses on work done to adapt REM to terrains with significant topography challenges. Inspired by the topographic feature of the RNM and using Rwanda as a case study, the author of \cite{drouin_geospatial_2018} describes a methodology to incorporate topography to the model. Rwanda serves as a great case-study for evaluating topographic-handling capabilities since its location at the East African Plateau means a lot of the landmass lies on challenging, hilly terrains. 
 
 The applications highlighted above demonstrate the various strengths of REM as a tool for large scale planning and the effectiveness of the underlying `bottom-up' clustering methodology. However, this paper proposes a different, novel computational approach to the integrated planning problem that has demonstrated great potential.  In contrast to the REM's ``bottom-up" method, we propose a ``top-down" approach, one which involves starting at the network level with all consumers connected to the grid before determining who may be better served by off-grid systems.
Thus rather than beginning at the individual consumer level, we start from this fully connected distribution network  and systematically partition the network, if necessary, to find less-expensive planning configurations. Given robust capabilities to design grid and off-grid distribution systems and estimate their associated costs, such a procedure may better capture the true cost elements that determine the overall electrification planning costs. Also, by identifying clusters from partitioning a reference distribution network designed by a robust network planning tool, we may better account for features and costs of geographic constraints such as topography or forbidden zones, which may inform the decision to designate areas for off-grid planning or otherwise. Another potential advantage that may arise from this approach is the ease of future transitioning of off-grid areas to grid extension regions. This is because all off-grid microgrid clusters are obtained by partitioning what corresponds to the future grid in which every consumer has been connected to the grid. Thus, the designated off-grid clusters may inherit information of the least future grid topology which may lead to lower future grid connection costs compared to other approaches.

Having introduced this procedure, in the next section, section IV., we discuss the computational details of our network partitioning algorithm.

\section{Top-down Methodology}
Herein, we present the details of the network partitioning computational methodology for identifying the least-cost electrification mode per consumer and consequently designing the recommended system. As previously mentioned, the overall idea is to first, design a reference distribution network for the region under consideration such that all consumers are connected to the existing grid, and subsequently greedily partition this network, using cost parameters, to identify those who are better connected to grid and otherwise. The designed network represents the ideal future network if all consumers were able to be connected to the grid and electrification costs were not a barrier. Since this distribution network is radial, all elements of the designed distribution network and their associated properties (such as costs) can then be represented in a tree data structure for computation purposes. This makes it possible to approach this electricity consumer bi-partitioning problem (into off-grid and grid-extension partitions) as a tree pruning problem and explore greedy tree-partitioning strategies to that effect.
\\
Thus, we assume that a network planner has a network design routine capable of designing an optimal (or quasi-optimal) reference network of all new consumers in a given region and quantifying the costs of every element in this distribution network. It should be noted that `Network Optimality' here refers the fact that the designed network from such a routine is expected to be the least-cost network.  Any reference network so designed can then be assumed to have the same topological characteristics of the future least-cost grid when all consumers become fully connected to the grid. Some examples of tools that can be used to generate this reference distribution network have been reviewed earlier. 
There are many properties of the distribution network that can be  extracted from a distribution network design. From the topological properties of the network, if we define pointers which show the hierarchy of power flow, from the grid all the way to the consumers and through every element in between, a tree data structure is obtained. In addition to pointers indicating hierarchy, other properties of elements in the network such as the length of line segments, the geo-location of equipment, the cost of all elements in the network, the voltage levels and other electrical, spatial and economic properties can also be adequately represented in the equivalent tree data structure of the network.

It is thus easy to see that for the equivalent tree of a designed radial distribution network, three main types of nodes will exist:

\begin{enumerate}
    \item Line Segment: These are the electrical distribution lines which distribute power to consumers or transmit from one element in the tree to another.   
    \item Consumers: Consumers are the load consumption nodes which ultimately must be fed and have their demands met. It is easy to see that for the equivalent tree of a radial distribution network, a node is a leaf node if and only if it is a consumer node.
    \item Transformers or sub-Stations: Transformer nodes are nodes which represent power transformer equipment for voltage transformation within the network. They link node elements  at different voltage levels together. 
\end{enumerate}

To understand this equivalent tree representation, consider the illustrative sample distribution network shown in Fig. 1 showing the physical network elements. 

\begin{figure}[h]
\centering
\includegraphics[scale = 0.3]{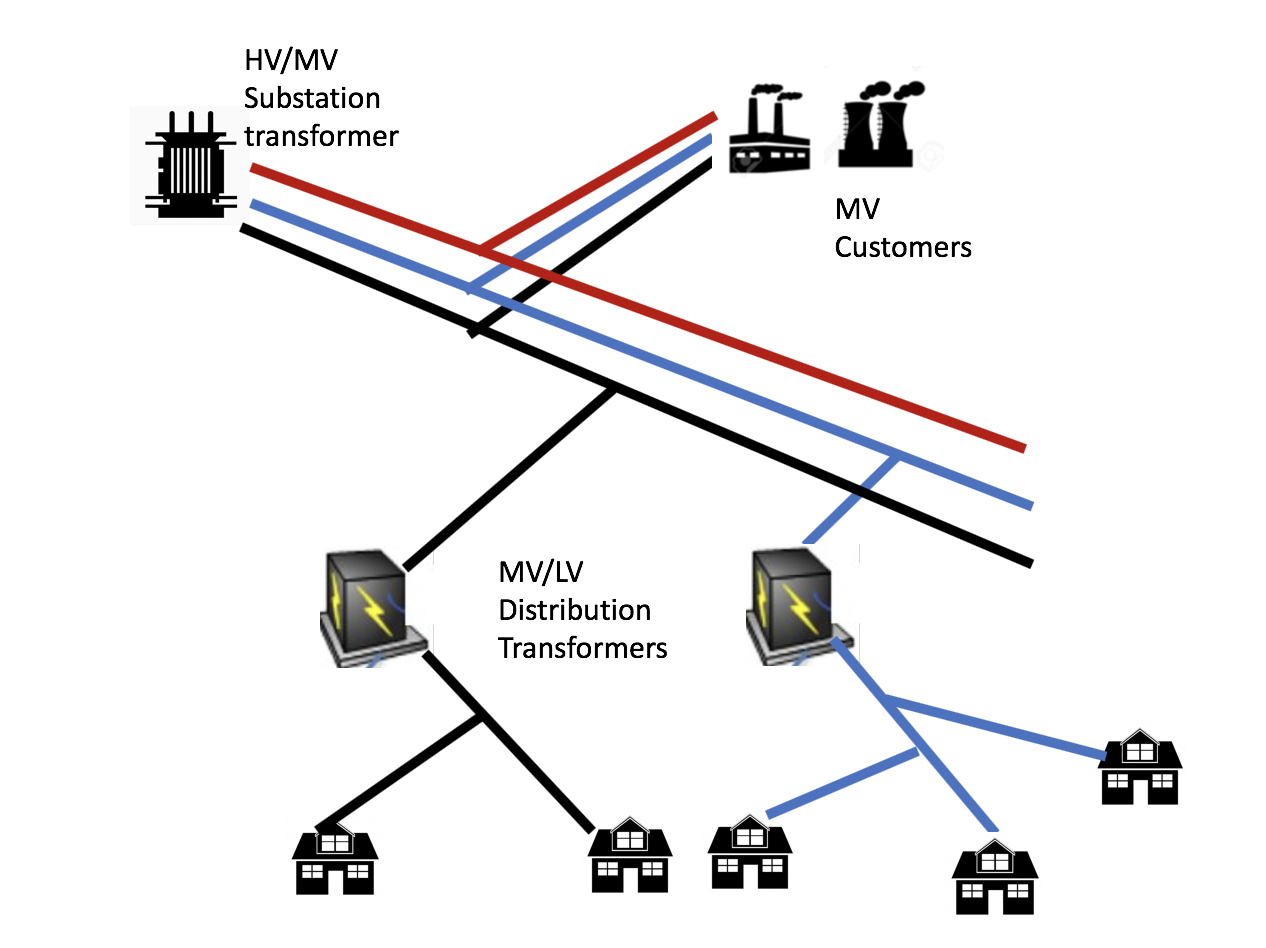}
\caption{Illustrative Physical Distribution Network}
\end{figure}

 Fig. 2 represents the equivalent tree for the radial distribution network, showing the nodes and terminologies which will be referred to in the rest of this paper. As seen, the physical elements of the network are all represented by equivalent nodes in the tree data structure, with each computational node also storing the properties of its associated physical network element such as its cost, and capacity. This equivalent tree representation of a distribution network can be extended to any radial distribution network, however large, rendering them amenable to computational procedures such as the partitioning-for-planning method proposed in this paper.
 
\begin{figure}[h]
\centering
\includegraphics[scale = 0.3]{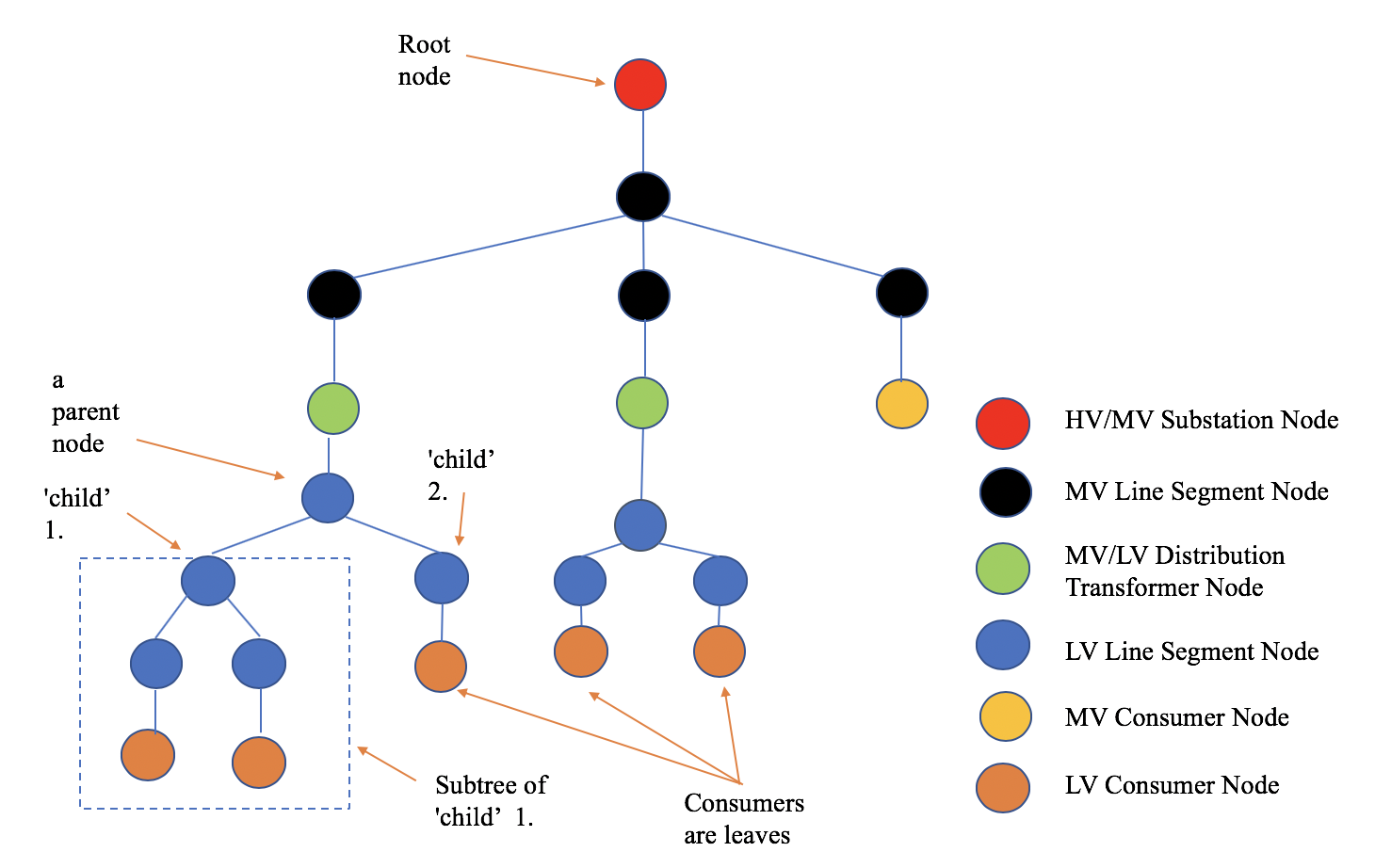}
\caption{Corresponding Tree Data Structure of Radial Distribution Network}
\end{figure}

\subsection{Greedy Tree Pruning}
Given our objective of partitioning or pruning the network tree to determine the least-cost electrification mode for every consumer node, a greedy approach to this problem may be considered. A greedy strategy means that the decision made locally at a node is the locally optimal one out of the two possible decisions i.e. to prune or otherwise. Pruning here is the computational equivalent of assigning consumers who would have otherwise being connected to the grid, to off-grid microgrid systems.  In deciding that a node (and the consumer nodes in its subtree) be pruned away and designated as off-grid consumers, or retained in the tree as part of the grid extension, we compare the costs with the benefits of the decision. The costs incorporated in REM and described in Section III, serve as a robust reference point for the system costs which we can consider in order to prune. As has been discussed in previous section, these costs are both the direct monetary costs i.e. the investment costs (distribution network of off-grid generation costs), the O and M costs, the direct energy and associated losses cost as well as the indirect societal cost of Non-Served Energy (CNSE) - the Cost of Non-Served Energy (a cost to penalize energy not provided due to less than 100\% reliability).

 At any node, the optimal pruning decision is thus one determined from a local evaluation of cost and benefit of retaining the consumers in the downstream subtree to the tree or grid.  For our proposed partitioning approach, we can evaluate these costs and benefit for every node in the network in order to inform any pruning decisions. For example, if a given node and its downstream nodes is to be pruned in the tree, the cost of introducing a new offgrid generation facility to electrify must be less than the savings in cost in the grid from having it removed. For every node $i \in N = \{1, 2,...n\}$ in an $n$-node tree, we can define and compute a decision value $\delta_i$ which tracks this decision as given by (1);

\begin{equation} \label{eq:1}
\delta_i = \gamma_i + \eta_i + \omega_i - \beta_i - \tau_i - \zeta_i - \sigma_i
\end{equation}

where  $\gamma_i$ and $\omega_i$  and $\eta_i$ are the costs incurred if the node's subtree is pruned. $\gamma_i$ is the downstream off-grid generation investment cost incurred if the node's subtree is pruned to off-grid, $\eta_i$ captures the net O \& M and management cost in providing off-grid generation rather than the grid while $\omega_i$ is the cost of non-served energy associated with the resulting offgrid consumers. $\gamma_i$  is also implemented such that it can capture the costs associated with the inclusion of an additional MV/LV transformer should the microgrid associated with the pruned subtree be a large (MV) microgrid.  
\\
$\beta_i$, $\tau_i$, $\zeta_i$ and $\sigma_i$ are the savings or benefits incurred if pruning occurs. $\beta_i$ is the total cost savings upstream if the downstream subtree is pruned, $\tau_i$ is the cost of non-served grid energy saved if the node's subtree is pruned  $\zeta_i$ is the grid energy cost saved after pruning, and $\sigma_i$ is the self cost of node $i$ which is saved if the node is pruned. $\sigma_i$  also captures costs associated with the electrical losses of the node's equivalent element.
A key assumption in our definition of $\delta_i$ is that there is negligible difference in network cost before and after pruning for the  downstream subtree consumers. Without this assumption, we would have to evaluate any gains in the overall network cost as a result of pruning and factor it in the computation of $\delta_i$. 
\\
If $\delta_i < 0,$ then pruning is the locally optimal decision and is undertaken.  An exhaustive search on all nodes in the tree, in a bottom-up fashion, can be undertaken until no further pruning is possible. Consumer nodes that are pruned can then be designated as off-grid customers while those still remaining in the tree are designated as grid-extension consumers. 
\\
In addition, the following criteria should also be met:
\begin{itemize}
    \item Bottom-up Traversal: The tree must be traversed in a bottom-up fashion. That is, the exploration begins from a leaf node and all nodes must be examined before their ancestors. 
    \item Pruning decisions are irreversible.
\end{itemize}

It is possible to traverse the tree in different ways such that the bottom-up criterion is satisfied.  While distance to the root seems a good criterion for tree traversal for our electrification problem, there are other important parameters, such as the magnitude of downstream demand, not captured by such a strategy. Our expectation of any good traversal strategy is that nodes which are farther from the grid and have higher downstream capacity  are  pruned earlier because they imply higher grid costs, and thus higher savings when pruned. Factoring both the distance to the grid with the power delivered in the traversal strategy also helps to capture the voltage drop, which is a very relevant cost driver in real-life distribution network planning.
\\ Considering this, we define a node property, $Moments$, $S$, which captures this combination of downstream power and distance to the grid. For a tree of nodes n, the $Moments$ $S_i$ at a node as:

   $S_i = S_i * L_i  + max\{S_{v1}, S_{v2}, S_{v3}\dots\}$  
   
   where $S_{v1}, S_{v2}, S_{v3}...$ are the moments of node $i$'s children and $L_i$ is the length of node $i$.

Thus, each node's $Moments$ value is at least as great as those of its downstream nodes . By quantifying both the downstream capacity and the distance to the grid via this parameter, we can prioritize nodes which when pruned will result in larger downstream capacities or relatively farther consumers  being disconnected from the grid.  The $Moments$ values can be initially pre-computed as the equivalent tree is constructed from the designed network.
Based on the above,  a procedure based on the combination of power-distance or $Moments$  can be described for identifying the next best candidate node to pruning evaluation as follows:
\begin{itemize}
        \item Pre-sort nodes from highest to lowest moments. 
    \item  Check sorted array if nodes have had all their downstream nodes evaluated or if they have no downstream node, if yes terminate tree traversal else go to next node in sort, proceed.
    \item Return next best candidate node for pruning as highest value in sorted nodes array.
\end{itemize}

\subsection{Overall Greedy Partitioning Algorithm}
By integrating the previously-described heuristics, an overall greedy partitioning procedure, for pruning the tree to identify the consumers to be connected to the grid and those to remain as off-grid candidates, can be described as in Algorithm 1.

\begin{algorithm}
\caption{Network Partitioning Algorithm}\label{euclid}
\begin{algorithmic}[0]
\Procedure{NetworkPartitioner}{}
   \State $tree\gets  \texttt{construct network tree procedure}$\Comment{tree has n nodes and all consumers initially grid-connected}
   \State $S_{offgrid}\gets \emptyset$
   \State $S_{total}\gets  \{1,2,...k \}$\Comment{There are k total consumers} 
   \While{\texttt{ unevaluated nodes are in $tree$}}

      \State $i\gets \texttt{get id of next node to evaluate}$
     \State $\delta_i = \gamma_i + \omega_i + \eta_i - \beta_i - \tau_i - \zeta_i - \sigma_i$
     \State \texttt{Set node $i$ as evaluated}
   
   \If{$\delta_i\leq 0$}
      \State $List_i\gets \texttt{get id of consumers in}$
      \State $\texttt{                subtree of $i$}$
      \State $S_{offgrid}\gets  S_{offgrid} \cup List_i $
      \State \texttt{Prune subtree of node $i$}
      \State \texttt{Update $tree$}
      
    \EndIf
    \EndWhile\label{euclidendwhile}
    \State $S_{ongrid}\gets S_{total} \setminus S_{offgrid}$
   \State \textbf{return} $S_{offgrid},S_{ongrid}$  
\EndProcedure
\end{algorithmic}
\end{algorithm}

In Algorithm 1, all consumers are initially connected to the grid-extension partition $S_{ongrid}$ while the offgrid partition $S_{offgrid}$ is empty. For every node $i$ visited in the tree, the local decision variable $\delta_i$ is computed and, if negative, the downstream consumer indices are appended to the offgrid partition $S_{offgrid}$. The tree properties are then updated and the process is repeated until all nodes have been evaluated. Recall that the downstream offgrid generation cost incurred is $\gamma_i$ while $\omega_i$ is the Cost of Non-Served Energy (CNSE) of node $i$'s downstream consumers if  offgrid. In addition, $\beta_i$, $\tau_i$, $\zeta_i$ and $\sigma_i$ are the upstream savings, the grid cost of non-served energy, the grid energy cost and the node self cost (the investment cost associated with a node) respectively.  The cost due to losses are also incorporated in the self cost when appropriate.

The downstream generation cost $\gamma_i$ is computed by simply calling a generation design function or look-up table using node $i$'s subtree consumer nodes properties as arguments. The look-up table function has already been previously incorporated in REM as described in \cite{Amatya_WP_2018} and \cite{ellman_reference_2015}. The grid cost of non served energy $\tau_i$ and grid energy cost, $\zeta_i$ are similarly determined by invoking these downstream consumer node data to functions which then compute them as documented in the agglomerative `bottom-up' clustering based version of REM \cite{Amatya_WP_2018}.
The upstream savings $\beta_i$ penalizes pruning that lead to little savings in cost of upstream grid infrastructure. To compute this, we evaluate the cost differences in all the ancestors to node $i$ before versus after pruning.

Note that it is important to update upstream tree properties of the tree after pruning to account for the expected changes in a distribution network. Some properties of other nodes in the tree may change due to the removal of downstream nodes.  Specifically, the following two properties need to be updated  with every tree pruning decision as they may also affect other node properties such as the $moments$ and $path-to-root$ which are used to identify next best pruning candidate.:
\begin{itemize}
        \item Power capacity: When nodes downstream are pruned, capacity of upstream ancestor nodes should  reduce correspondingly.  
    \item Self Cost: This is typically a function of the power capacity. The self cost also incorporates the cost due to losses when appropriate.
\end{itemize}

Executing Algorithm 1 returns the partitions of consumers designated as off-grid and grid-extension consumers. The grid-extension consumer partition can then be passed to a network design module in order to obtain final grid-extension network designs, one of the objectives of the rural electrification planning process. For those consumers designated as off-grid, we can then run an additional off-grid microgrid clustering procedure to cluster these off-grid consumers - based on total cost - into different microgrid clusters before then designing their associated off-grid networks. 

\section{Results}
As previously mentioned, an objective of this work is to explore how top-down based approaches to the consumer electrification mode recommendation and clustering problem compare to bottom-up methods such as that previously implemented REM `bottom-up' method described in  \cite{ellman_reference_2015}. 
To this end, the computational procedure previously described was implemented and tested on a number of cases in order to understand and comparatively evaluate its performance.  A test case (with 6688 consumers) was selected to study the performances of these two approaches. The test case was selected given its unique characteristics such as the varying geospatial densities of consumers in different parts of the region as well as the demand characteristics of the region. Drawn from a real-life planning case, the test case requires the incorporation of multiple demand profiles instead of a much simpler case of assuming a homogenous residential demand profile for all consumers. 

Fig. 3 shows the geographic distribution of these 6688 consumers in the test case region as well as the surrounding existing grid. We now present results obtained when REM's bottom-up method and the partitioning procedure are applied. MATLAB R2018a was used as the programming platform for implementing the top-down method and performing the analyses.

\begin{figure}[h]
\centering
\includegraphics[scale=0.5]{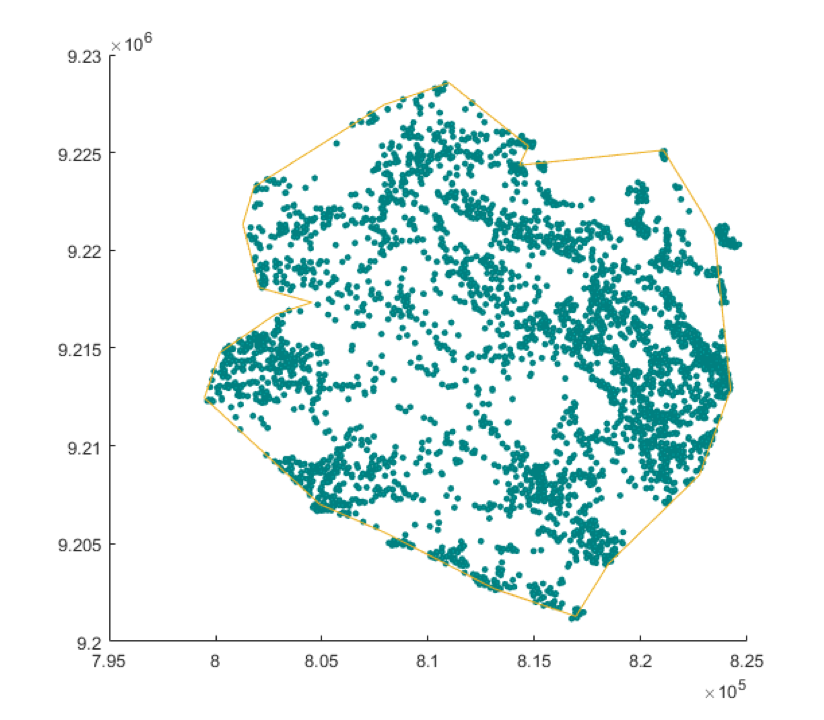}
\caption{Consumers Distribution for Test case}
\end{figure}

For the top-down method, $NetworkPartitioner$ was utilized as the tree partitioning procedure as presented in Algorithm 1. The DT-agglomerative or `bottom-up' method is the same presented in \cite{Amatya_WP_2018}. 

\subsection{Base Case Results}
Using the same input data from the generation and network equipment catalog, as based on the expertise of the MIT-Comillas Universal Energy Access Lab; authors of \cite{Amatya_WP_2018} , both approaches were applied on the test case. Before undertaking sensitivity analysis (the variation of input parameters to observe changes in input) on the test case, a reference base case has to be defined. The sensitive parameters under consideration in the application presented are the input fuel cost and reliability of the existing grid network. For the reference case, the fuel cost is set at $\$0.8/L$ and the grid reliability at $90\%$. The grid energy cost value is also set at $\$0.08/kWh$. Running the test case using these values lead to the results presented in Fig. 4 and Fig. 5 and summarized in Table I and Table II.
As can be observed, the results for this particular scenario and its associated input data show the top-down partitioning method leading to lower costs. The fraction of consumers assigned to off-grid versus grid-extension are also similar. The next subsections show how this test case responds to variations in certain parameters and data values.

\begin{table}[h!]
\caption{ Base Case Bottom-up Results Summary}
\begin{tabu} to 0.5\textwidth { | X[2,l] | X[c] |X[c] |X[c] | X[r] | }
\hline
 System Type & Microgrids &  Isolated & Grid  & All \\
 \hline
 \hline
 Number of Customers  & 0 & 145 & 6543 & 6688\\
 \hline
Annual System Cost (\$) & 0 & 139,706 & 3,378,329 & 3,518,035 \\
 \hline
\end{tabu}
\end{table}

\begin{table}[h!]
\caption{ Base Case Top-down Results Summary}
\begin{tabu} to 0.5\textwidth { | X[2,l] | X[c] |X[c] |X[c] | X[r] | }
\hline
 System Type & Microgrids &  Isolated & Grid  & All \\
 \hline
 \hline
 Number of Customers & 0 & 175 & 6513 & 6688\\
 \hline
 Annual System Cost  (\$) & 0 & 115,756 & 3,330,305 & 3,446,061 \\
 \hline
\end{tabu}
\end{table}

\begin{figure}[h]
\raggedleft
\includegraphics[scale = 0.6]{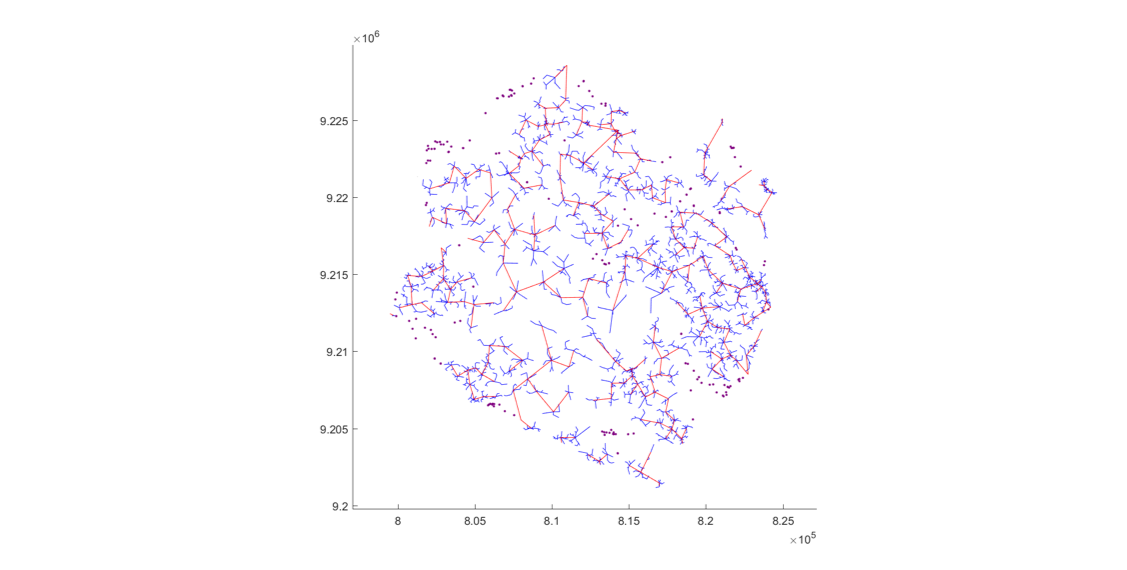}
\includegraphics[scale=0.6]{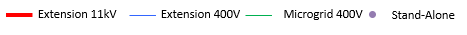}
\caption{Base Case: Test case Result using Agglomerative `$Bottom-up$' based method }
\end{figure}

 \begin{figure}[h]
\includegraphics[scale=0.6]{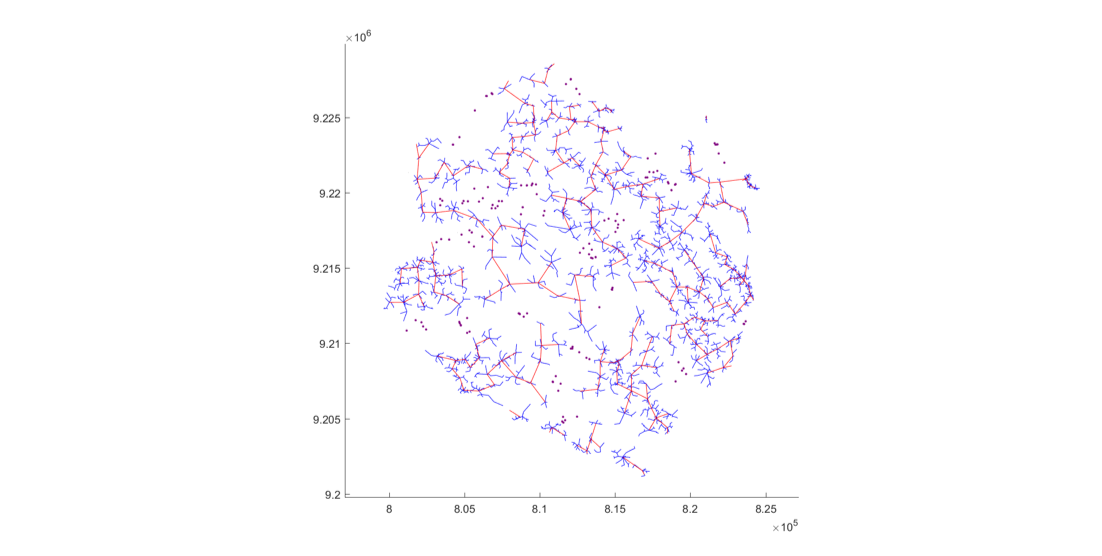}
\includegraphics[scale=0.6]{images/legend.PNG}
\caption{Base Case: Test case Result using `Top-down' method}
\end{figure}

\clearpage
\newpage
\mbox{}

\subsection{Sensitivity Analysis: Fuel Cost}
Keeping all other data and parameters constant, the fuel cost was varied and results examined to see differences and sensitivities of both approaches to this parameter. Since the fuel cost parameter affects primarily the off-grid generation cost, intuitively, higher fuel cost should lead to less microgrid clusters than otherwise. 
Tables V and VI summarize the fuel cost sensitivity analysis results for both approaches.
\\

\begin{table}[h!]
\caption{ Fuel Cost Sensitivity Results Summary using Bottom-up Clustering Method}
\begin{tabu} to 0.5\textwidth { | X[l] |X[c] |X[c] |X[c] |X[c] |X[c] | X[r] | }
\hline
 \multicolumn{3}{|c|}{Fuel Cost = \$0.7/L } & \multicolumn{2}{|c|}{Fuel Cost = \$0.6/L } & \multicolumn{2}{|c|}{Fuel Cost = \$0.5/L} \\
\hline
 System Type & Number of Customers &  Annual System Cost (\$) & Number of Customers &  Annual System Cost (\$) & Number of Customers &  Annual System Cost (\$) \\
 \hline
 \hline
 Microgrids  & 0 & 0 &  3602 & 1,707,699  & 6233 & 2,930,615 \\
 \hline
 Isolated  & 157 & 125,336 & 234 & 186,807 & 158 & 126,134 \\
  \hline
 Grid  & 6333 & 3,304,488 & 2852 & 1,574,481 & 297 & 159,990 \\
  \hline
  All  & 6688 & 3,520,583 & 6688 & 3,455,689 & 6688 & 3,216,740 \\
  \hline
\end{tabu}
\end{table}

\begin{table}[h!]
\caption{ Fuel Cost Sensitivity Results Summary using Top-down Partitioning Method}
\begin{tabu} to 0.5\textwidth { | X[l] |X[c] |X[c] |X[c] |X[c] |X[c] | X[r] | }
\hline
 \multicolumn{3}{|c|}{Fuel Cost = \$0.7/L } & \multicolumn{2}{|c|}{Fuel Cost = \$0.6/L } & \multicolumn{2}{|c|}{Fuel Cost = \$0.5/L} \\
\hline
 System Type & Number of Customers &  Annual System Cost (\$) & Number of Customers &  Annual System Cost (\$) & Number of Customers &  Annual System Cost (\$) \\
 \hline
 \hline
 Microgrids  & 0 & 0 &  3960 & 1,872,419  & 5685 & 2,656,906 \\
 \hline
 Isolated  & 161 & 128,529 & 529 & 422,311 & 170 & 135,714 \\
  \hline
 Grid  & 6527 & 3,327,160 & 3199 &  1,295,432 & 833 & 490,731 \\
  \hline
  All  & 6688 & 3,455,689 & 6688 & 3,500,162 & 6688 & 3,283,352 \\
  \hline
\end{tabu}
\end{table}

The results from the off-grid generation input fuel cost sensitivity analysis follow the expected intuitive trends; decreasing the off-grid generation diesel fuel cost leads to more off-grid consumer candidates. In terms of cost, both the top-down and bottom-up clustering approaches lead to similar system costs values; with the top-down leading to slightly lower system costs in all but one of the scenarios presented in the tables. In addition, in Fig. 6 \& Fig. 7, we present the GIS representations of the final recommended grid-extension and off-grid system designs for the case of Diesel Cost = \$0.5/Liter. As can be observed, results from both approaches also have significant, though not exact, overlap in electrification mode recommendations, despite resulting from two different underlying computational approaches.

\begin{figure}[h]
\centering
\includegraphics[scale=0.6]{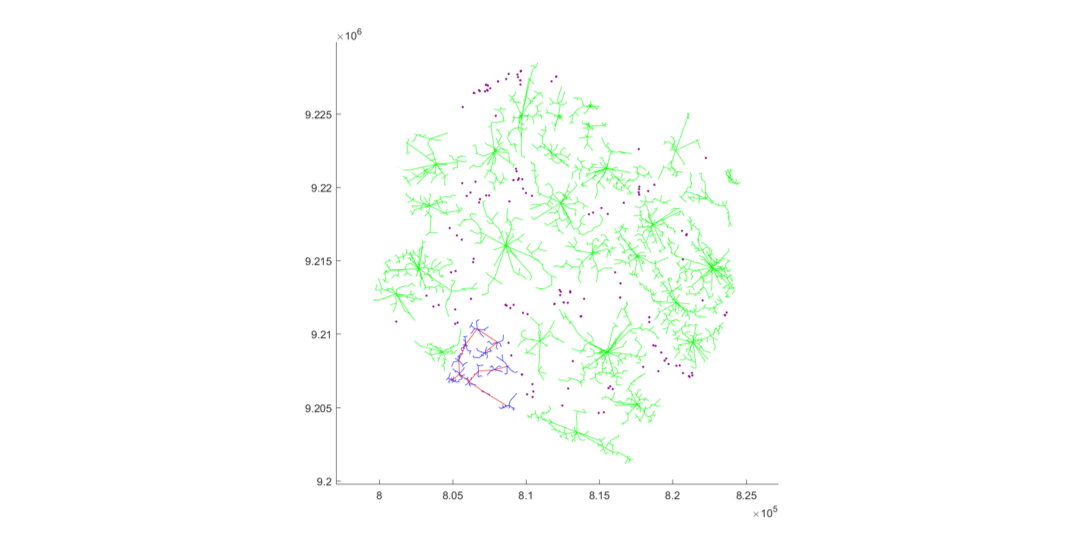}
\includegraphics[scale=0.6]{images/legend.PNG}
\caption{Fuel Cost = \$0.5/L: Test case Result using Agglomerative `$Bottom-up$' based method }
\end{figure}

 \begin{figure}[h]
\centering
\includegraphics[scale=0.6]{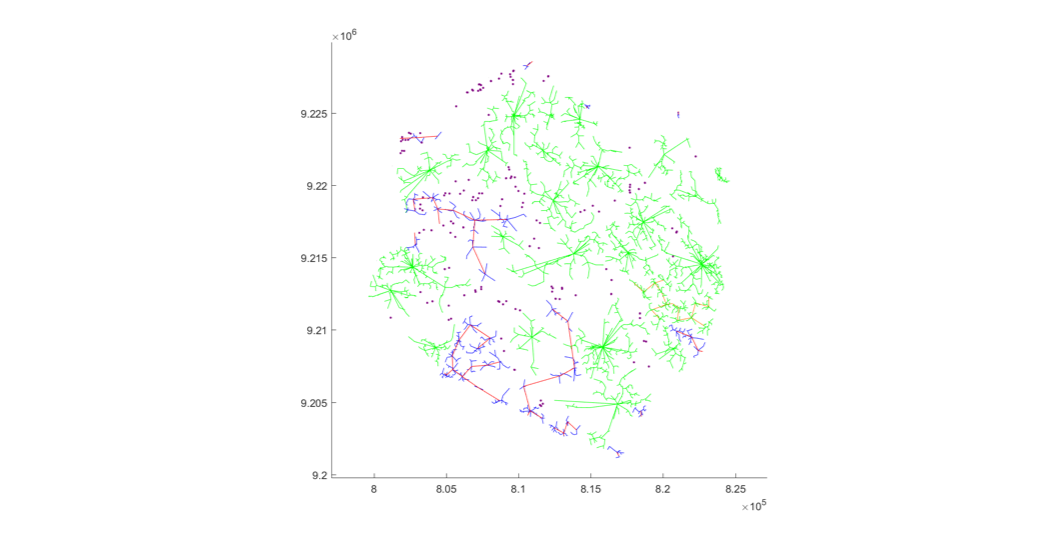}
\includegraphics[scale=0.6]{images/legend.PNG}
\caption{Fuel Cost = \$0.5/L: Test case Result using using `Top-down' method}
\end{figure}

\subsection{Sensitivity Analysis: Grid Reliability}
The sensitivity analysis was also repeated using the grid reliability level as the varying parameter. Intuitively, the expectation would be that at higher grid reliability, there would be more consumers assigned to the grid extension and vice-versa.
Tables V and VI summarize the grid reliability sensitivity analysis results for both approaches.
\\
\begin{table}[h!]
\tiny
\caption{ Grid Reliability Sensitivity Results Summary using Bottom-up Clustering Method}
\begin{tabu} to 0.5\textwidth { | X[l] |X[c] |X[c] |X[c] |X[c] |X[c] | X[r] | }
\hline
 \multicolumn{3}{|c|}{Grid Reliability = 100\% } & \multicolumn{2}{|c|}{Grid Reliability = 90\% } & \multicolumn{2}{|c|}{Grid Reliability = 70\%} \\
\hline
 System Type & Number of Customers &  Annual System Cost (\$) & Number of Customers &  Annual System Cost (\$) & Number of Customers &  Annual System Cost (\$) \\
 \hline
 \hline
 Microgrids  & 0 & 0 &  0 & 0  & 6233 & 0,0 \\
 \hline
 Isolated  & 34 & 27,143 & 145 & 139,706 & 161 & 128,529 \\
  \hline
 Grid  & 6654 & 1,909,549 & 6543 & 3,378,329 & 6527 & 3,327,160 \\
  \hline
  All  & 6688 & 1,936,691 & 6688 & 3,518,035 & 6688 & 3,455,689 \\
  \hline
\end{tabu}
\end{table}

\begin{table}[h!]
\tiny
\caption{ Grid Reliability Sensitivity Results Summary using Top-down Partitioning Method}
\begin{tabu} to 0.5\textwidth { | X[l] |X[c] |X[c] |X[c] |X[c] |X[c] | X[r] | }
\hline
 \multicolumn{3}{|c|}{Grid Reliability = 100\% } & \multicolumn{2}{|c|}{Grid Reliability = 90\% } & \multicolumn{2}{|c|}{Grid Reliability = 70\%} \\
\hline
 System Type & Number of Customers &  Annual System Cost (\$) & Number of Customers &  Annual System Cost (\$) & Number of Customers &  Annual System Cost (\$) \\
 \hline
 \hline
 Microgrids  & 0 & 0 &  0 & 0 & 198 & 90,759 \\
 \hline
 Isolated  & 37 & 29,538 & 175 & 115,756 & 157 & 125,336 \\
  \hline
 Grid  & 6651 & 1,829,370 & 6513 &  3,330,305 & 6333 & 3,304,488 \\
  \hline
  All  & 6688 & 1,858,908 & 6688 & 3,446,061 & 6688 & 3,520,583 \\
  \hline
\end{tabu}
\end{table}

As with the fuel cost sensitivity results, the results from the grid reliability sensitivity analysis follow the expected intuitive trends: decreasing the grid reliability led to more off-grid consumer candidates.  Results from both approaches also have significant overlap in recommendations with similar cost values. At high reliability, the top-down approach led to lower system costs with the bottom-up dominating at lower reliability for the test case.

\section{Conclusion and Further Developments}
This paper presents a novel partitioning-based computational method to address an important aspect of planning electricity infrastructure for those without access; the identification of the least-cost electrification mode for every consumer in a region under analysis. Implementing the proposed methodology, as we have shown, led to similar electrification cost solutions when bench-marked with a `bottom-up' method incorporated in an already existing planning model, REM. The sensitivity analyses showed that the `top-down' approach led to lower system cost values than the `bottom-up' method at relatively higher grid reliability and fuel cost values and the opposite was also true. The proposed method of partitioning the network also allows factors which may be otherwise difficult to model, such as savings upstream of the grid due to planning decisions, to be captured.  Upstream network reinforcement costs - including upstream impact all the way to transmission and generation levels -  may also be better incorporated using the approach presented and this can serve as an area of future research work.

\ifCLASSOPTIONcaptionsoff
  \newpage
\fi

\bibliographystyle{IEEEtran}
\bibliography{references}

\end{document}